\documentclass[aps,pra,reprint,superscriptaddress,floatfix,preprintnumbers,showkeys]{revtex4-1}

\usepackage{graphicx}
\usepackage{color}
\usepackage{amssymb}
\usepackage{amsmath}
\usepackage{epsfig}
\usepackage{bm}
\usepackage[american]{babel}
\usepackage{hyperref}
\usepackage{braket}
\usepackage[T1]{fontenc}
\usepackage{txfonts}

\DeclareGraphicsExtensions{.pdf,.png,.jpg}
\hypersetup{colorlinks=true,urlcolor=blue,breaklinks=true,citecolor=blue,linkcolor=blue,pdfstartview=FitH,pdfpagemode=UseNone}

\newcommand{\inda}{a}
\newcommand{\indb}{b}
\newcommand{\indc}{c}
\newcommand{\intmeasure}{\mathrm{d}^3 r}   
\newcommand{\ud}{\mathrm{d}}
\newcommand{\ve}[1]{\mathbf{#1}}                 
\newcommand{\velocity}{\ve{v}}                
\newcommand{\vorticity}{\ve{\Omega}}    
\newcommand{\sftensor}{Q}                              
\newcommand{\spin}{s}                                       
\newcommand{\mz}{m_F}                                    
\newcommand{\qpuv}{\hat{\ve{B}}_{\mathrm{q}}}   
\newcommand{\nemvec}{\hat{\ve{d}}}      
\newcommand{\averagespin}{\langle |\ve{\spin}| \rangle} 
\newcommand{\zsub}{{\mathrm{Z}}}                                         
\newcommand{\cartsub}{\mathrm{C}}                                     
\newcommand{\transpose}{\mathrm{T}}                

\begin{document}
\title{Evolution of an isolated monopole in a spin-1 Bose--Einstein condensate}
\date{November 16, 2016}
\author{Konstantin Tiurev}\email{konstantin.tiurev@aalto.fi}
\affiliation{QCD Labs, COMP Centre of Excellence, Department of Applied Physics, Aalto University,
P.O. Box 13500, FI-00076 Aalto, Finland}
\author{Pekko Kuopanportti}
\affiliation{School of Physics and Astronomy, Monash University, Victoria 3800, Australia}
\author{Andr\'as M\'arton Gunyh\'o}
\affiliation{QCD Labs, COMP Centre of Excellence, Department of Applied Physics, Aalto University,
P.O. Box 13500, FI-00076 Aalto, Finland}
\author{Masahito Ueda} 
\affiliation{Department of Physics, University of Tokyo, Hongo 7-3-1, Bunkyo-ku, Tokyo 113-0033, Japan}
\affiliation{RIKEN Center for Emergent Matter Science (CEMS), Wako, Saitama 351-0198, Japan}
\author{Mikko M\"ott\"onen} 
\affiliation{QCD Labs, COMP Centre of Excellence, Department of Applied Physics, Aalto University,
P.O. Box 13500, FI-00076 Aalto, Finland}
\affiliation{Department of Mathematical Information Technology, University of Jyv{\"a}skyl{\"a}, P.O. Box 35, FI-40014 Jyv{\"a}skyl{\"a}, Finland}

\begin{abstract}
We simulate the decay dynamics of an isolated monopole defect in the nematic vector of a spin-1 Bose--Einstein condensate during the polar-to-ferromagnetic phase transition of the system. Importantly, the decay of the monopole occurs in the absence of external magnetic fields and is driven principally by the dynamical instability due to the ferromagnetic spin-exchange interactions. An initial isolated monopole is observed to relax into a polar-core spin vortex, thus demonstrating the spontaneous transformation of a point defect of the polar order parameter manifold to a line defect of the ferromagnetic manifold. We also investigate the dynamics of an isolated monopole pierced by a quantum vortex line with winding number $\kappa$. It is shown to decay into a coreless Anderson--Toulouse vortex if $\kappa=1$ and into a singular vortex with an empty core if $\kappa=2$. In both cases, the resulting vortex is also encircled by a polar-core vortex ring.
\end{abstract}
\preprint{DOI: \href{http://dx.doi.org/10.1103/PhysRevA.94.053616}{10.1103/PhysRevA.94.053616}}

\maketitle

\section{\label{sc:intro}Introduction}

Topological defects are ubiquitous to many areas of physics, such as condensed matter, cosmology, and exactly solvable models~\cite{Mer1979.RMP51.591,Nak2003_book,Kib1976.1387, Cruz2007, Kitaev2006}. Experimentally, an ideal platform to create and observe them in quantum matter is provided by gaseous Bose--Einstein condensates (BECs)~\cite{And1995.Sci269.198,Kaw2012.PhysRep520.253}, which allow accurate experimental control over many characteristic parameters and enable direct imaging of the quantum-mechanical order parameter field. The variety of available topological defects is especially rich in the case of BECs with spin degrees of freedom due to their many possible order parameter manifolds and underlying symmetries~\cite{Ued2014.RepProgPhys77.122401}. Presently, the experimentally realized topological structures in BECs are diverse:
singly and multiply quantized vortices~\cite{Lovegrove.PRA.93.033633,Mat1999.PRL83.2498,Mad2000.PRL84.806,Abo2001.Sci292.476,Lea2002.PRL89.190403,Iso2007.PRL99.200403};
solitons and vortex rings~\cite{Science.Denschlag.2000,And2001.PRL86.2926}; coreless~\cite{Lea2003.PRL90.140403,Les2009.PRL103.250401}, polar-core~\cite{Sad2006.Nat443.312}, and solitonic~\cite{Don2014.PRL113.065302} vortices; skyrmions~\cite{Cho2012.PRL108.035301,Cho2012.NJP14.053013}; quantum knots~\cite{Hal2016.NatPhys12.478}; and monopoles~\cite{Ray2014.Nat505.657,Ray2015.Sci348.544}.

Recently, isolated monopoles were experimentally observed in the polar manifold of a three-dimensional $^{87}$Rb spin-1 BEC by Ray \emph{et al.}~\cite{Ray2015.Sci348.544}. The existence of such topological point defects in the nematic vector of the condensate is permitted because the second homotopy group $\pi_2$ for the polar order parameter space $G_\mathrm{p}=[S^2\times \mathrm{U}(1)]/\mathbb{Z}_2$~\cite{Zho2003.IntJModPhysB17.2643,Ued2014.RepProgPhys77.122401} is nontrivial. Namely, it is isomorphic to the additive group of integers: $\pi_2(G_\mathrm{p})\cong\mathbb{Z}$.  In contrast, the opposing ferromagnetic phase of a spin-1 BEC forbids genuine point defects since its order parameter space $G_\mathrm{f}=\mathrm{SO}(3)$ yields $\pi_2(G_\mathrm{f})\cong\emptyset$~\cite{Mer1979.RMP51.591}.  It does, however, permit the so-called Dirac monopole configuration~\cite{Sav2003.PRA68.043604,Pie2009.PRL103.030401,Ruo2011.PRA84.063627}, which exhibits a radial monopole field in the superfluid vorticity $\vorticity$ and serves as a simulation of a charged quantum particle interacting with a classical magnetic point charge, i.e., the scenario first considered by Dirac in his seminal theoretical work~\cite{Dir1931.PRSLA133.60}. As experimentally verified~\cite{Ray2014.Nat505.657}, this kind of monopole induces in the BEC order parameter a vortex filament, described by Dirac as a nodal line, that extends from the location of the vorticity monopole to the surface of the condensate. Although this configuration is topologically equivalent to the defect-free ground state, it is energetically and dynamically reminiscent of a vortex line, except for the endpoint. In fact, the ferromagnetic manifold $G_\mathrm{f}$ also supports a topologically nontrivial class of vortices, as evidenced by its nontrivial first homotopy group $\pi_1(G_\mathrm{f})\cong \mathbb{Z}_2$. Since the polar and ferromagnetic components can simultaneously exist in different regions of a single inhomogeneous spin-1 BEC, the cores of these nontrivial vortices tend to be filled with the polar component, prompting the term \emph{polar-core vortex}~\cite{Iso2001.JPSJ70.1604,Miz2002.PRL89.030401,Mar2002.PRA66.053604,Miz2002.PRA66.053610,Miz2004.PRA70.043613,Bul2003.PRL90.200401}.

In Ref.~\cite{Tiurev2015.pra.93.033638}, the eventual decay of the isolated polar-phase monopole into the Dirac monopole configuration was studied numerically under conditions similar to their first experimental realization~\cite{Ray2015.Sci348.544}. However, the dynamics were exclusively investigated in the presence of a quadrupole magnetic field, and consequently the linear Zeeman coupling to the atomic spins was found to steer the formation of the Dirac monopole configuration. Other effects such as the spin-exchange interactions or the three-body recombination were reported to have a negligible effect on the dynamics. Quantitatively, these results are well aligned with the previous findings that (i) the isolated monopole~\cite{Ruo2003.PRL91.190402} and indeed the entire polar phase of the $^{87}$Rb spin-1 BEC~\cite{Sad2006.Nat443.312} are expected to be unstable at low magnetic fields and (ii) the local strong-field seeking state in the quadrupole magnetic field, i.e., the  ferromagnetic position-dependent spin state that minimizes the Zeeman energy, gives rise to a vorticity monopole $\vorticity \propto \hat{\ve{r}}/r^2$~\cite{Sav2003.PRA68.043604,Pie2009.PRL103.030401,Ruo2011.PRA84.063627}. Essentially, the latter fact also forms the basis for the method~\cite{Pie2009.PRL103.030401} used to create and observe the Dirac monopole configuration in Ref.~\cite{Ray2014.Nat505.657}. Zeeman steering of the atomic spins with multipole magnetic fields can also be utilized to topologically imprint~\cite{Lea2002.PRL89.190403,Nak2000.PhysicaB284.17,Iso2000.PRA61.063610,Oga2002.PRA66.013617,Mot2002.JPCM14.13481} and pump~\cite{Mot2007.PRL99.250406,Xu2008.PRA78.043606,Xu2008.NJP11.055019,Xu2010.PRA81.053619,Kuo2010.JLTP161.561,Kuo2013.PRA87.033623} vortices into BECs.

This paper investigates a scenario that is conceptually different from the one in Ref.~\cite{Tiurev2015.pra.93.033638}; namely, we study the behavior and ultimate fate of an isolated monopole defect in a dynamical polar-to-ferromagnetic quantum phase transition in the absence of any magnetic fields. This transition is driven by the ferromagnetic nature of the atomic spin-exchange interactions and results in mixing of polar and ferromagnetic phases discussed in Refs.~\cite{Lovegrove.PRA.93.033633,Oh.PRL.112.160402}. The initially flow-free quadrupole nematic state is shown to transform into a polar-core spin vortex. If, on the other hand, the monopole is accompanied by a singly or a doubly quantized $\mathrm{U}(1)$ vortex, the configuration is observed to decay, respectively, into a coreless Anderson--Toulouse vortex or a singular spin vortex with an empty core; both types of vortices are additionally encircled by a polar-core vortex ring. Importantly, all the observed quantum phase transitions are robust in the sense that including or excluding dissipation in the form of three-body recombinations causes no qualitative changes in the decay dynamics. 

The remainder of this article is organized as follows. In Sec.~\ref{sc:theory}, we outline the mean-field theory of spin-1 BECs and characterize the order parameter manifolds. Section~\ref{sc:methods} describes our simulation methods, the results of which are presented and analyzed in Sec.~\ref{sc:results}. Finally, Sec.~\ref{sc:discussion} concludes the paper with a brief discussion.

\section{\label{sc:theory}Theory}

\subsection{Equation of motion for a spin-1 condensate}

The mean-field order parameter of a spin-1 BEC can be expressed as $\bm{\Psi}(\ve{r})=\sqrt{n(\ve{r})}\bm{\xi}(\ve{r})$, where $n(\ve{r})$ is the number density of atoms in the condensate and $\bm{\xi}(\ve{r})\in\mathbb{C}^3$ is a three-component spinor that satisfies $\bm{\xi}^\dagger \bm{\xi} = 1$. The time evolution of the order parameter at sufficiently low temperatures is accurately described by the spin-1 Gross--Pitaevskii equation~\cite{Ho1998.PRL81.742,Ohm1998.JPSJ67.1822}
\begin{equation}\label{eq:GP}
\begin{aligned}
i\hbar \frac{\partial}{\partial t}\bm{\Psi}(\ve{r}) &= \Big[ -\frac{\hbar^2}{2m}\nabla^2+V(\ve{r}) + g_{F} \mu_\mathrm{B} \ve{B}(\ve{r},t) \cdot \ve{F}  \\&
 - i\, \Gamma n^{2}(\ve{r})  + g_\mathrm{d} n(\ve{r}) + g_\mathrm{s} n(\ve{r}) \ve{\spin}(\ve{r})\cdot \ve{F} \Big] \bm{\Psi}(\ve{r}),
\end{aligned}
\end{equation}
where  $m$ is the mass of the atoms, $V(\ve{r})$ is an external optical trapping potential, $\Gamma$ is the three-body recombination rate, $\ve{\spin}=\bm{\xi}^\dagger \ve{F} \bm{\xi}$ is the local average spin, and $\ve{F}$ is a vector of the dimensionless spin-1 matrices satisfying $\left[F_\inda,F_\indb\right]=i \sum_\indc \varepsilon_{\inda\indb\indc}F_\indc$; here $\varepsilon_{\inda\indb\indc}$ is the Levi-Civita symbol and $\inda,\indb,\indc \in \left\{x,y,z\right\}$. The coupling constants characterizing the local density--density and the local spin-exchange interactions are given by $g_\mathrm{d}=4\pi\hbar^2(a_0+2a_2)/3m$ and $g_\mathrm{s}=4\pi\hbar^2(a_2-a_0)/3m$, respectively, where $a_f$ is the $s$-wave scattering length corresponding to the scattering channel with total two-atom hyperfine spin $f$. The optical trap is assumed to be harmonic, $V(\ve{r})=m[\omega_r^2(x^2+y^2)+\omega_z^2 z^2]/2$, where $\omega_r$ and $\omega_z$ are the radial and axial trapping frequencies, respectively.  The linear Zeeman term $g_{F} \mu_\mathrm{B} \ve{B}(\ve{r},t) \cdot \ve{F}\,\bm{\Psi}$ couples the condensate atoms to the external magnetic field $\ve{B}$. Here $g_F$ is the Land\'{e} factor and $\mu_\mathrm{B}$ is the Bohr magneton. The quadratic Zeeman effect is observed to be negligible~\cite{Tiurev2015.pra.93.033638} and therefore is not included in Eq.~\eqref{eq:GP}. In this work, we consider both the spherically symmetric case $\omega_r = \omega_z$ and the experimentally realized scenario with $\omega_z/\omega_r \approx 1.32$~\cite{Ray2015.Sci348.544}.

For $\Gamma=0$ and $\ve{B}=\ve{0}$, Eq.~\eqref{eq:GP} conserves the total number of condensate particles $N=\int n\left(\ve{r}\right) \intmeasure$, the total energy $E$, the total magnetization  $\ve{M}=\int n\left(\ve{r}\right)  \ve{\spin}\left(\ve{r}\right) \intmeasure$~\cite{note_magnetization_conservation}, and the $z$ component of the orbital angular momentum, $L_z=-i\hbar \int  \bm{\Psi}^\dagger\left(\ve{r}\right)\hat{\ve{z}}\cdot \ve{r}\times \nabla\bm{\Psi} \left(\ve{r}\right) \intmeasure$. Although we mostly consider here dissipative dynamics with $\Gamma > 0$, the resulting states are similar to those arising from unitary dynamics for the relatively small, realistic value of $\Gamma$ we employ. 

In our analysis, it is convenient to utilize two different bases for the spin degree of freedom. The first is the so-called Cartesian basis~\cite{Ohm1998.JPSJ67.1822,Mue2004.PRA69.033606}, in which the spin matrices are given by $\left( F_\indc \right)_{\inda\indb}=-i\varepsilon_{\inda \indb \indc}$ and the spinor $\bm{\Psi}=\left(\Psi_x,\Psi_y,\Psi_z\right)^\transpose_\cartsub$ transforms as an ordinary vector under spin rotations. The second is the eigenbasis of $F_z$, which is obtained from the first by the unitary transformation
\begin{equation}
\begin{pmatrix} \Psi_{+1} \\ \Psi_0 \\ \Psi_{-1} \end{pmatrix}_\zsub = \frac{1}{\sqrt{2}}
\begin{pmatrix}
-\Psi_{x} + i\Psi_{y}\\
\sqrt{2}\Psi_{z} \\
\Psi_{x} + i\Psi_{y} 
\end{pmatrix}_\zsub
\end{equation}
and is convenient for describing states that are symmetric about the $z$ axis. Above and in what follows, column vectors carry a subscript $\cartsub$ or $\zsub$ to indicate the employed basis.

\subsection{Quadratic tensor: nematic and spin ordering}

With the help of the Cartesian basis, we can express the local spin as the vector product $\ve{\spin}=\ve{l}\times \ve{m}$, where $\ve{l},\ve{m}\in\mathbb{R}^3$ are given by the decomposition $\sqrt{2} \xi_\inda = l_\inda + i m_\inda$, where $\inda \in \left\{x,y,z\right\}$. For $|\ve{\spin}|<1$, we define the nematic vector  $\nemvec \in\mathbb{R}^3$ as a unit-length eigenvector corresponding to the largest eigenvalue of the real symmetric unit-trace $3\times 3$ matrix
\begin{equation}\label{eq:sftensor}
\begin{aligned}
\sftensor_{\inda\indb}\left(\ve{r}\right) &= \delta_{\inda\indb} - \frac{1}{2} \bm{\xi}^{\dagger} \left(F_\inda F_\indb + F_\indb F_\inda\right) \bm{\xi} \\ &=\frac{1}{2}\left(\xi_\inda \xi_\indb^\ast + \xi_\indb \xi_\inda^{\ast} \right)=\frac{1}{2} \left(l_\inda l_\indb + m_\inda m_\indb \right),
\end{aligned}
\end{equation}
which describes spin fluctuations~\cite{Mue2004.PRA69.033606}. In the \textit{pure polar} phase with $\ve{\spin}=\textbf{0}$, we must have  $\ve{l} \parallel \ve{m}$; it then follows from Eq.~\eqref{eq:sftensor} that the polar-phase order parameter can be written in the Cartesian basis as $\ve{\Psi}(\ve{r})=\sqrt{n(\ve{r})}\exp\left[i\varphi(\ve{r})\right]\nemvec (\ve{r})$ for some $\varphi \in\mathbb{R}\ (\textrm{mod}\ 2\pi)$~\cite{note_nematic_order}.

The \textit{pure ferromagnetic} phase with $|\ve{\spin}|=1$ is the other extreme case. Here the triad $\left(\ve{\spin},\ve{l},\ve{m}\right)$ forms an orthonormal basis of $\mathbb{R}^3$. Since the order parameter space $G_\mathrm{f}=\mathrm{SO}(3)$ does not support point defects, the decay of an isolated monopole into a ferromagnetic state has to result in a topologically different type of structure. Indeed, as we show in Sec.~\ref{sc:results}, the isolated monopole actually decays into a spin vortex associated with a nontrivial element of $\pi_1\left(G_\mathrm{f}\right)$. We also point out that since the eigenvalues of $\sftensor$ in descending order are $\lambda_1=1/2+\sqrt{{1-|\ve{\spin}|^2}}/2$, $\lambda_2=1/2-\sqrt{{1-|\ve{\spin}|^2}}/2$, and  $\lambda_3=0$, the pure ferromagnetic phase has $\lambda_1=\lambda_2=1/2$ and, consequently, an ill-defined nematic vector $\nemvec$.

\subsection{Winding-number and symmetry considerations}

To gain some preliminary insight into the monopole decay, let us consider a spinor that belongs to the polar manifold and exhibits a hedgehog monopole, $\bm{\xi}_{\textrm{h}}=(\xi_x,\xi_y,\xi_z)^{\transpose}_\cartsub= \exp(i\kappa \phi)\hat{\ve{r}}$, where we also allow for the existence of a straight singular vortex along the $z$ axis with the winding number $\kappa\in\mathbb{Z}$. In terms of the matrices $F_\inda$, this spinor can also be constructed as $\exp(i\kappa \phi) \hat{\ve{r}} = \exp(i\kappa \phi) \exp(-i\phi F_z) \exp(-i\theta F_y)\hat{\ve{z}}$, where $\left(\theta,\phi\right)$ are the spherical coordinates. In the $z$-quantized basis, it becomes
\begin{equation}\label{eq:polar_psi}
\bm{\xi}_{\textrm{h}}
=\frac{e^{i \kappa \phi}}{\sqrt{2}}\begin{pmatrix}
-e^{-i \phi} \sin\theta\\
\sqrt{2} \cos\theta\\
e^{i \phi} \sin\theta
\end{pmatrix}_\zsub.
\end{equation}
Thus, the componentwise winding numbers about the $z$ axis are $\kappa -1$, $\kappa$, and $\kappa + 1$  in the components $\mz = 1$, $0$, and $-1$, respectively. Irrespective of the value of $\kappa$, this state yields the nematic vector $\ve{\hat{d}}=\pm \ve{\hat{r}}$, except at the $z$ axis where $\ve{\hat{d}}$ is not defined if $\kappa\neq 0$. 

As observed in Sec.~\ref{sc:results}, the componentwise winding numbers about the $z$ axis tend to be conserved during the specific dynamics that we study. Therefore, it is relevant to ask what the componentwise winding numbers $\kappa -1$, $\kappa$, and $\kappa + 1$ correspond to in the ferromagnetic manifold. It turns out that this depends strongly on the underlying rotational symmetry of the resulting spin texture. With this in mind, we construct the purely ferromagnetic Cartesian-basis spinor $\bm{\xi}_{\textrm{fm}}=-\exp(-i\phi F_z) \exp[-i\beta(\ve{r}) F_y]\exp(i\kappa \phi F_z) \left(\hat{\ve{x}}+i\hat{\ve{y}}\right)/\sqrt{2}$, where $\beta:\mathbb{R}^3 \mapsto \left[0,\pi\right]$ is a smooth function. In the $z$-quantized basis, we obtain
\begin{equation}\label{eq:ferromagnetic_psi}
\bm{\xi}_{\textrm{fm}}=
e^{i \kappa \phi}\begin{pmatrix}
e^{-i \phi} \cos^2\frac{\beta\left(\ve{r} \right) }{2}\\
\sqrt{2} \cos\frac{\beta\left(\ve{r} \right) }{2} \sin\frac{\beta\left(\ve{r} \right) }{2}\\
e^{i \phi}  \sin^2\frac{\beta\left(\ve{r} \right)}{2}
\end{pmatrix}_\zsub,
\end{equation}
where componentwise winding numbers about the $z$ axis are identical to those in Eq.~\eqref{eq:polar_psi}. On the one hand, for $\beta\left(\ve{r}\right)=\theta$, the spin texture is spherically symmetric, $\ve{\spin}=\hat{\ve{r}}$, and Eq.~\eqref{eq:ferromagnetic_psi} corresponds to the so-called Dirac monopole configuration; for $\kappa = 1$ ($\kappa=-1$), there is a singular vortex half-line at $z\leq 0$ ($z\geq 0$), whereas for $\kappa \neq \pm 1$, a singular vortex line extends along the entire $z$ axis. On the other hand, for cylindrically symmetric $\beta\left(\ve{r}\right)=\tilde{\beta}\left(\sqrt{x^2+y^2}\right)$, Eq.~\eqref{eq:ferromagnetic_psi} describes a mass or a spin vortex along the $z$ axis, the exact nature of which depends on both the function $\tilde{\beta}$ and the value of $\kappa$.  For example, by setting $\tilde{\beta}\equiv \pi/2$ and $\kappa$ even, Eq.~\eqref{eq:ferromagnetic_psi} corresponds to a singular, topologically nontrivial spin vortex, which tends to morph into the polar-core vortex in more realistic situations where parts of the BEC reside in the polar phase. Our simulations in Sec.~\ref{sc:results} indicate that the initial spherical-like symmetry of the isolated monopole is not preserved during its evolution under Eq.~\eqref{eq:GP}: The defect decays into a spin vortex instead of a Dirac monopole configuration that might be expected from the spherical symmetry.

\section{\label{sc:methods}Methods}

The original method to create the isolated monopole configuration in a polar-phase condensate is described in detail in Ref.~\cite{Ray2015.Sci348.544}; we only outline it here. The $^{87}$Rb condensate is initially prepared in the spin state $\ket{F,\mz}=\ket{1,0}$ in an external magnetic field $\ve{B}(\ve{r},t) = \ve{B}_\mathrm{q}(\ve{r})  + \ve{B}_\mathrm{b}(t)$, where $\ve{B}_\mathrm{b}(t) = B_\mathrm{b}(t) \ve{\hat{z}}$ is a spatially homogeneous bias field and $\ve{B}_\mathrm{q}(\ve{r}) = B_\mathrm{q}'(x \ve{\hat{x}} + y \ve{\hat{y}}-2z \ve{\hat{z}})$ is a quadrupole magnetic field. First, the  gradient $B_\mathrm{q}'$ of the quadrupole field is linearly ramped from zero to $3.7$ \textrm{G/cm} at bias field strength $B_\mathrm{b} = 1$ \textrm{G}. The bias field is then ramped to zero in two stages: the fast 10-ms ramp to $10$ \textrm{mG} and the subsequent adiabatic creation ramp to zero at the rate $\ud B_\mathrm{b}/\ud t  = - 0.25\textrm{ G/s}$. Ideally, this results in a vanishing superfluid velocity and a monopole state with the nematic vector $\hat{\ve{d}} = \hat{\ve{B}}_{\textrm{q}}$, where $\hat{\ve{B}}_{\textrm{q}}=\ve{B}_\mathrm{q}/| \ve{B}_\mathrm{q}|$ is the unit vector of the quadrupole field. Immediately after the creation ramp has concluded at $t=0$, we instantaneously switch off the quadrupole field and simulate the subsequent in-trap dynamics with $\ve{B} = \ve{0}$. 

In the simulations, the particle number is initially $N=2.1\times 10^5$, and the optical trapping frequencies are $\omega_r =2\pi\times 124$~Hz and $\omega_z=2\pi\times 164$~Hz, as in the experiments of Ref.~\cite{Ray2015.Sci348.544}. The atom-loss parameter due to three-body recombinations is set to $\Gamma_0=\hbar\times 2.9\times 10^{-30}\textrm{ cm}^6/\textrm{s}$~\cite{Bur1997.PRL79.337,Sta1998.PRL80.2027} throughout the paper. We set the $s$-wave scattering lengths to the literature values for $^{87}$Rb, $a_0=5.387$~nm and $a_2=5.313$~nm, which renders the spin-exchange interactions weakly ferromagnetic with $g_\mathrm{s}= g_{\mathrm{s}0} :=-0.00462\times g_\mathrm{d}$~\cite{Kem2002.PRL88.093201}. However, to better elucidate their role in the phase transition, we also investigate cases where $g_\mathrm{s}$ is instantaneously ramped at $t=0$ from $g_{\mathrm{s}0}$ to a smaller value for $t>0$, corresponding to more strongly ferromagnetic condensates. Furthermore, we note that the spin-exchange interactions do not play a particularly important role for $t<0$ due to the presence of the magnetic field and that the state at $t=0$ resides within the polar manifold to a reasonable approximation; therefore, we can also interpret $t=0$ as the moment when the spin-exchange interactions are quenched from antiferromagnetic  ($g_\mathrm{s} > 0$) to ferromagnetic ($g_\mathrm{s} < 0$), with the postquench dynamics subsequently observed. 

Prior to simulating the monopole creation process, we find the polar-state order parameter in the initial strong uniform magnetic field by using the successive over-relaxation algorithm~\cite{Teukolsky}. The subsequent dynamics are explored according to Eq.~\eqref{eq:GP} with the help of an operator-splitting method~\cite{Teukolsky}, fast Fourier transformations, and a time step of $2 \times 10^{-4} / \omega_r$. The simulated region is a cube of volume $\left(24\, a_{r}\right)^3$, where $a_r = \sqrt{\hbar /m\omega_{r}} = 1.02~\mu$m. We use 200 grid points per dimension. 

In the figures below, we apply, at will, homogeneous rotations to the spin fields for improved visibility of the resulting vortex structures. These rotations, however, do not affect the topology or the energy density.

\section{\label{sc:results}Results}

\begin{figure}[t]	
\includegraphics[width=0.95\columnwidth]{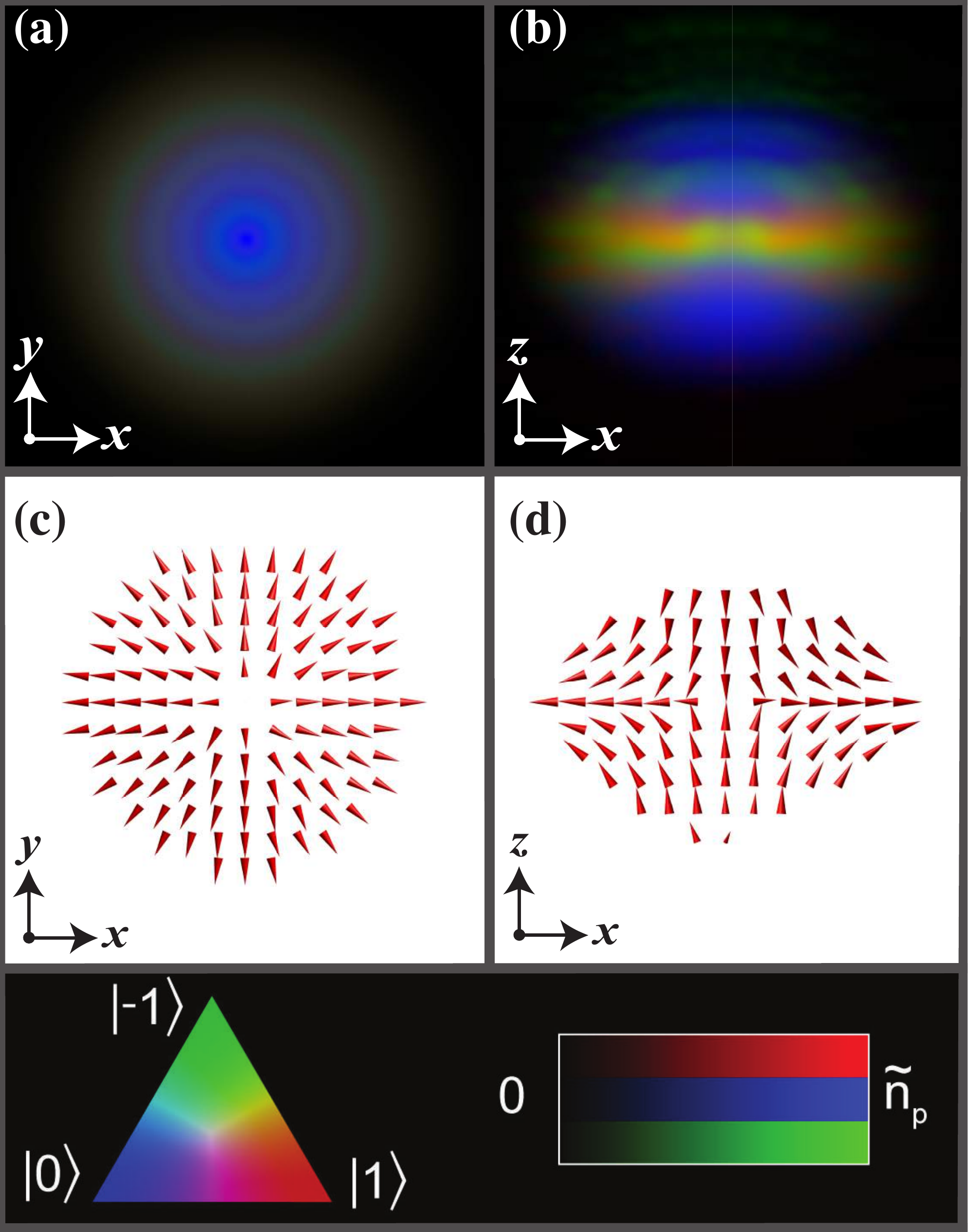}
\caption{\label{fig:1}Isolated monopole in the polar phase of a spin-1 BEC immediately after the simulated creation process. The spin-componentwise particle densities are integrated over (a)~$z$ and (b)~$y$. Different colors correspond to particles in different $z$-quantized spin states  $\ket{\mz}$, with the color and intensity scales given in the bottom panel; the peak column density is $\tilde{n}_{\textrm{p}} = 2.8 \times 10^{11} \textrm{ cm}^{-2}$. The configuration of the nematic vector $\nemvec$ is illustrated in the planes (c)~$z=0$ and (d)~$y=0$. The field of view is $15.5 \times 15.5\ {\mu}\textrm{m}^2$ in each panel.}
\end{figure}

Figure~\ref{fig:1} shows the state of the BEC immediately after the simulated monopole creation process has concluded, i.e., at $t=0$. The resulting configuration of the nematic vector field is in good agreement with $\nemvec =\qpuv$. However, due to nonadiabatic effects and spin-exchange collisions, the expectation value of the local spin magnitude $\averagespin(t)=  \int n(\ve{r},t)|\ve{\spin}(\ve{r},t)| \intmeasure/N(t)$ is about $0.2$ already at the end of the creation process. After the magnetic field is instantaneously switched off at $t=0$, the polar-to-ferromagnetic phase transition takes place, with the spin-exchange interaction energy
\begin{equation}
E_{\mathrm{s}} = - \frac{1}{2} |g_{\textrm{s}}|  \int n^2 \ve{\spin}^2 \intmeasure \leq 0
\end{equation}
decreasing from its unfavorably high value at $t=0$, converting into the kinetic, potential, and density--density-interaction energy of the BEC, and dissipating away due to the three-body recombinations.  Unless otherwise specified, we use the parameter values corresponding to the actual monopole creation experiments~\cite{Ray2014.Nat505.657,Ray2015.Sci348.544}, with $g_\mathrm{s}=g_{\mathrm{s}0}$ and $\Gamma=\Gamma_0$. 

\begin{figure}[t]
\includegraphics[width=1.00\columnwidth]{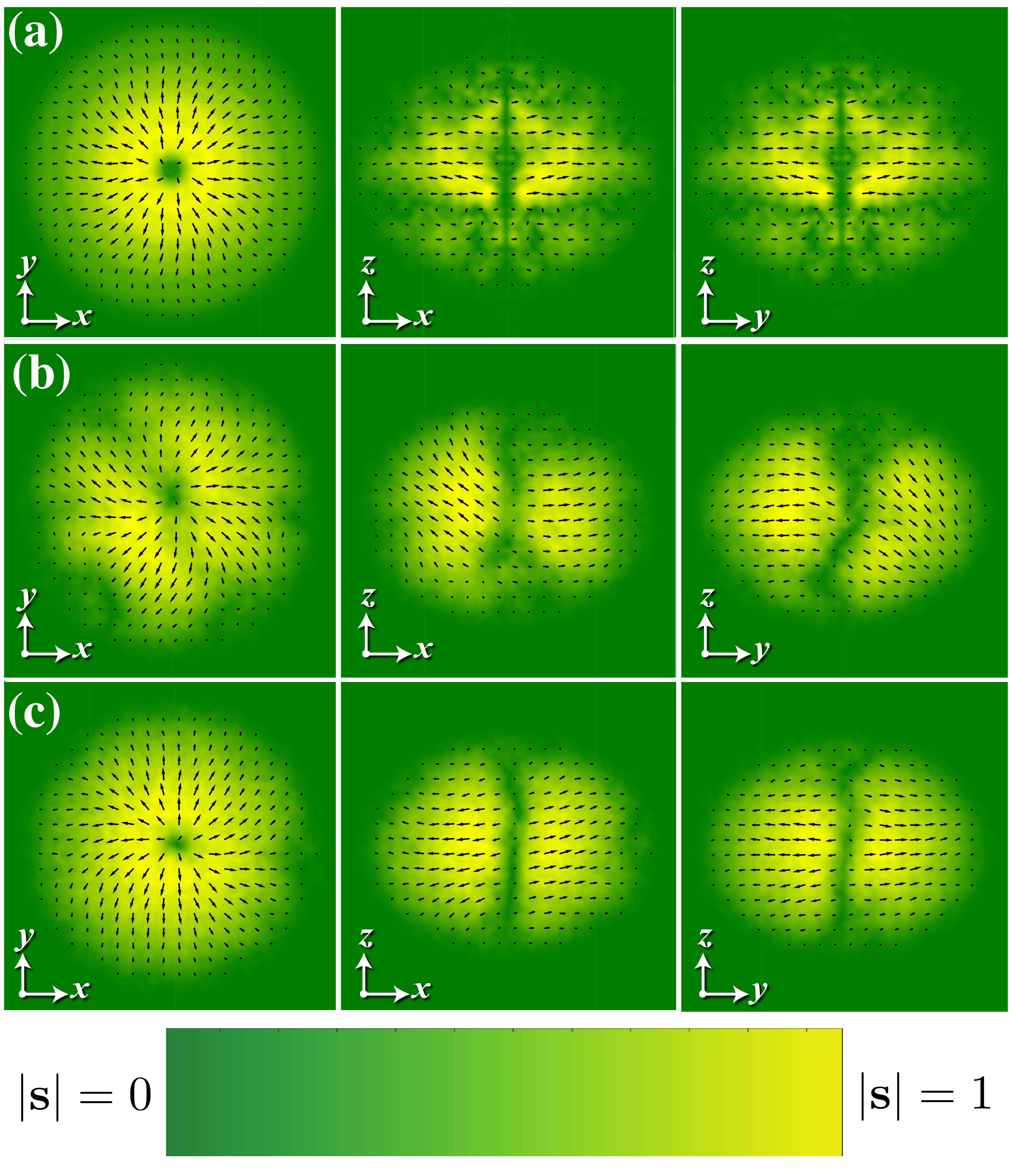}
\caption{\label{fig:2}Decay of the isolated monopole shown in Fig.~\ref{fig:1}. Each panel presents by arrows the projection of the local spin vector $\ve{\spin}(\ve{r})$ onto the shown coordinate plane, with the background color representing the magnitude $|\ve{\spin}(\ve{r})|$. Rows (a) and (b) show the resulting spin textures after (a) 100~ms and (b) 750~ms of temporal evolution with the spin-exchange interaction strength $g_\mathrm{s}=g_{\mathrm{s}0}:=-0.00462\times g_\mathrm{d}$ that corresponds to ferromagnetic spin-1 \textsuperscript{87}Rb. Row (c) is for 750~ms of waiting time with an enhanced interaction strength $g_\textrm{s} = 4 g_{\textrm{s0}}$. The field of view is $15.5 \times 15.5\ {\mu}\textrm{m}^2$ in each panel.}
\end{figure}

Figure~\ref{fig:2} depicts the spin field $\ve{\spin}$ well after the monopole creation. At $t=100$ ms [Fig.~\ref{fig:2}(a)],  the isolated monopole defect in the polar phase has transformed into a dominantly ferromagnetic state containing a polar-core vortex, the axial symmetry of which is observed to be broken at $t=750$ ms~[Fig.~\ref{fig:2}(b)]. The polar-core-vortex configuration is observed to be stable during the whole time interval studied. Although the vortex filament precesses in the cloud, it tends to be aligned with the $z$ axis, thus minimizing its length. An identical simulation with much stronger ferromagnetic interactions ($g_\mathrm{s} = 4 g_{\mathrm{s}0}$) exhibits similar qualitative behavior but with stronger localization of the vortex [Fig~\ref{fig:2}(c)]. Similar final states also emerge if spatially uncorrelated random noise with an amplitude of <1\% is applied to the state at $t=0$ (data not shown).

\begin{figure}[h]
\includegraphics[width=0.95\columnwidth]{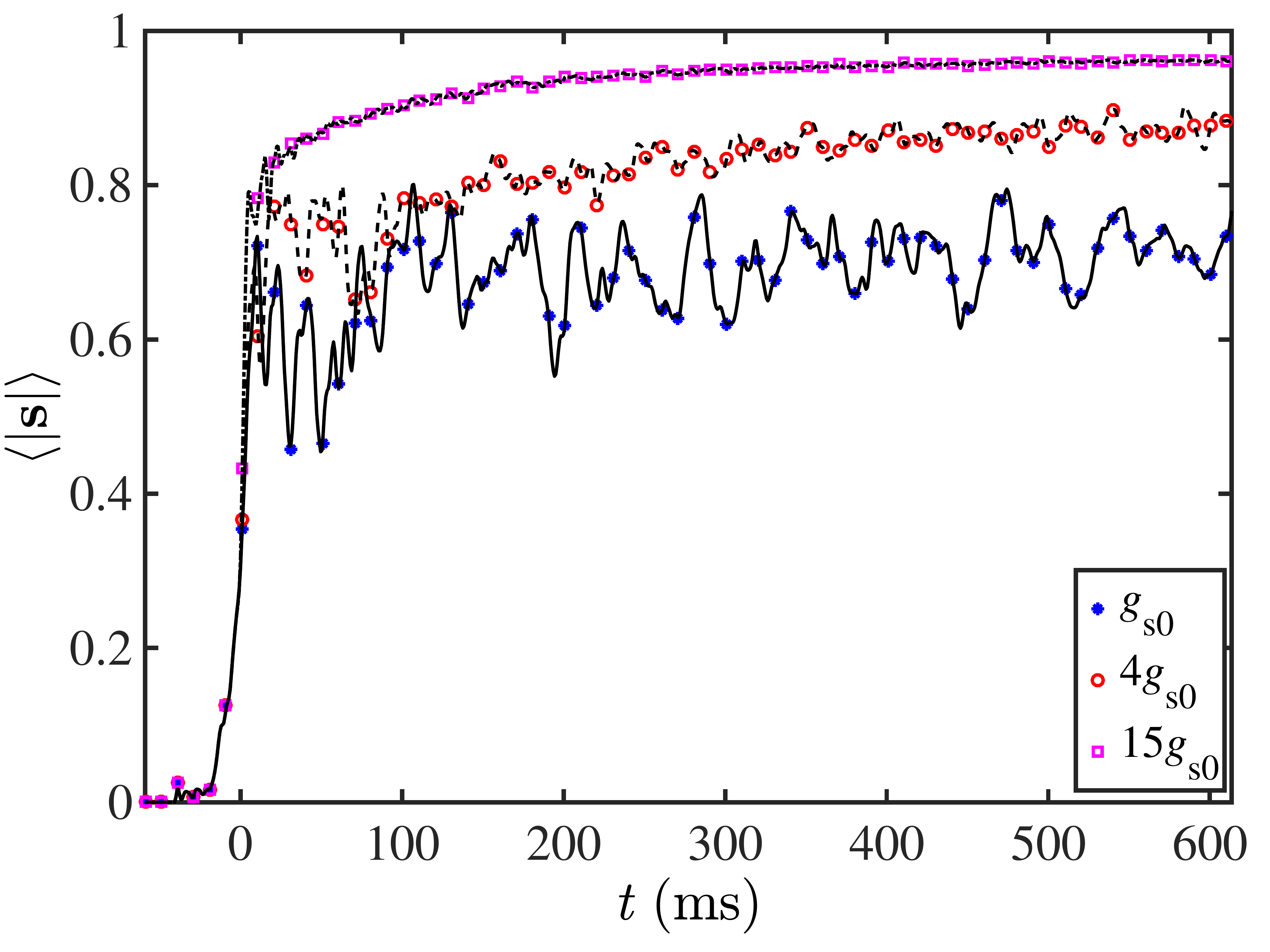}
\caption{\label{fig:3}Temporal evolution of the local spin magnitude $\averagespin (t)$ for three different spin-exchange interaction strengths $g_\mathrm{s}$. Here $g_{\mathrm{s}0}=-0.00462\times g_\mathrm{d}$ corresponds to the actual spin-1 \textsuperscript{87}Rb. At $t=0$, the BEC is in the state shown in Fig.~\ref{fig:1} in all three cases. The curves are cubic splines to guide the eye.}
\end{figure}

The temporal evolution of the local spin magnitude $\averagespin(t)$ is shown in Fig.~\ref{fig:3} for three different postquench values of $g_{\textrm{s}}$. In the simulation corresponding to the natural spin-exchange interaction strength $g_{\textrm{s}}= g_{\textrm{s}0}$, $\averagespin$ oscillates around 0.7, and thus a significant amount of the condensate still resides in the polar phase at $t=600\textrm{ ms}$. With increasing postquench $|g_{\textrm{s}}|$, this equilibrium value increases and the amplitude of the oscillations around it decreases; for $g_\mathrm{s}=15\,g_{\textrm{s}0}$, $\averagespin$ seems to saturate to~$0.9$. On the other hand, the spin magnitude is observed to  initially increase exponentially: $\averagespin(t)\propto \exp\left(t/\tau\right)$, where $\tau\approx 10\ \textrm{ms}$ is practically independent of the magnitude of $g_{\textrm{s}}$. 

\begin{figure}[ht]
\includegraphics[width=0.95\columnwidth]{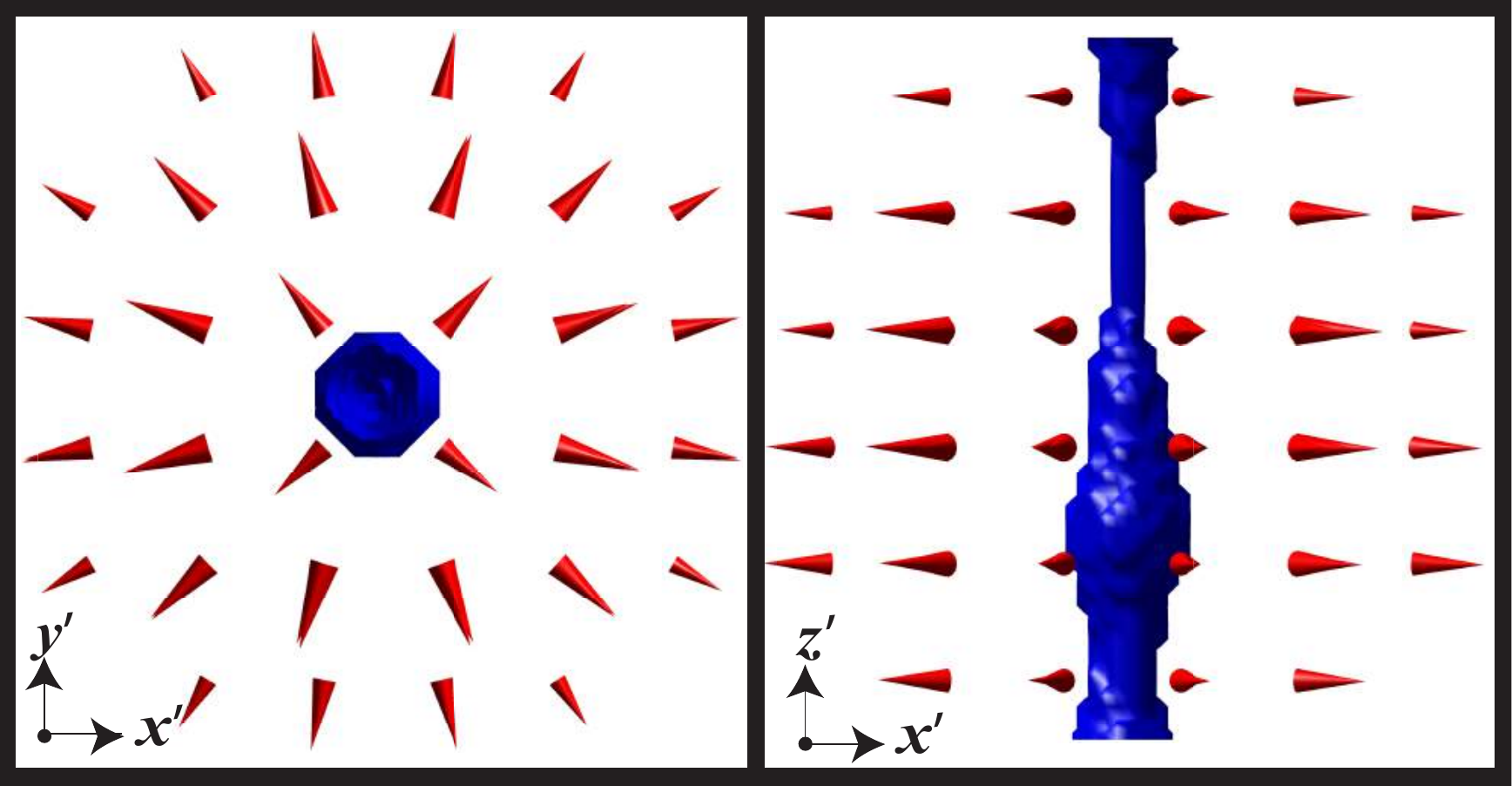}
\caption{\label{fig:4}Polar-core spin vortex obtained after 200 ms of decay of an ideal monopole configuration $\nemvec(\ve{r}) = \hat{\ve{r}}$. The red cones show the local spin field $\ve{\spin}(\ve{r})$, with their size proportional to $|\ve{\spin}(\ve{r})|$.  The blue isosurface corresponding to $|\ve{\spin}|=0.1$ illustrates the depleted spin density along the vortex core. The field of view is $8 \times 8\ {\mu}\textrm{m}^2$ in both panels, and the simulation employs the natural spin-exchange interaction strength $g_\mathrm{s}=g_{\mathrm{s}0}$. The primed coordinate system is obtained by rotating the original coordinate system such that  $\hat{\mathbf{z}}'$ is pointing along the vortex core.}
\end{figure}

The simulations presented in Figs.~\ref{fig:1}--\ref{fig:3} all have $\omega_z/\omega_r \approx 1.32$ and start with the  experimental creation process~\cite{Ray2015.Sci348.544}, resulting in the monopole state of Fig.~\ref{fig:1} with $\nemvec(\ve{r}) \approx \qpuv$ at $t=0$. In order to demonstrate that qualitatively our results are not specific to this initial state, let us also study the case of an ideal monopole configuration in a spherically symmetric optical trap with $\omega_z=\omega_r$. To this end, we begin the simulation from an exact polar-phase BEC with $\nemvec =\hat{\ve{r}}$ produced by fixing the spinor components according to Eq.~\eqref{eq:polar_psi} at $t=0$. As in the case of the experimental parameters, the isolated monopole again decays into a polar-core spin vortex (Fig.~\ref{fig:4}). This simulation clearly demonstrates that the spherical symmetry of the initial nematic state breaks down when the vortex emerges. We can also conclude that the nematic textures $\nemvec = \hat{\ve{r}}$ and $\nemvec=\qpuv$, which are topologically equivalent and hence correspond to the same singly quantized point defect associated with the second homotopy group $\pi_2(G_\mathrm{p})\cong\mathbb{Z}$, both decay into a state containing the same line-defect type, the polar-core spin vortex, associated with the nontrivial first homotopy group $\pi_1(G_\mathrm{f})\cong\mathbb{Z}_2$ of the ferromagnetic manifold. A simulation starting from an exact polar-phase BEC with $\nemvec=\qpuv$ also yields a polar-core spin vortex similar to Fig.~\ref{fig:4} (data not shown). All cases discussed above show qualitatively similar dynamics even if the three-body recombination is excluded from the model by setting $\Gamma = 0$ in Eq.~\eqref{eq:GP} (data not shown).

\begin{figure}[h!]
\includegraphics[width=0.886\columnwidth]{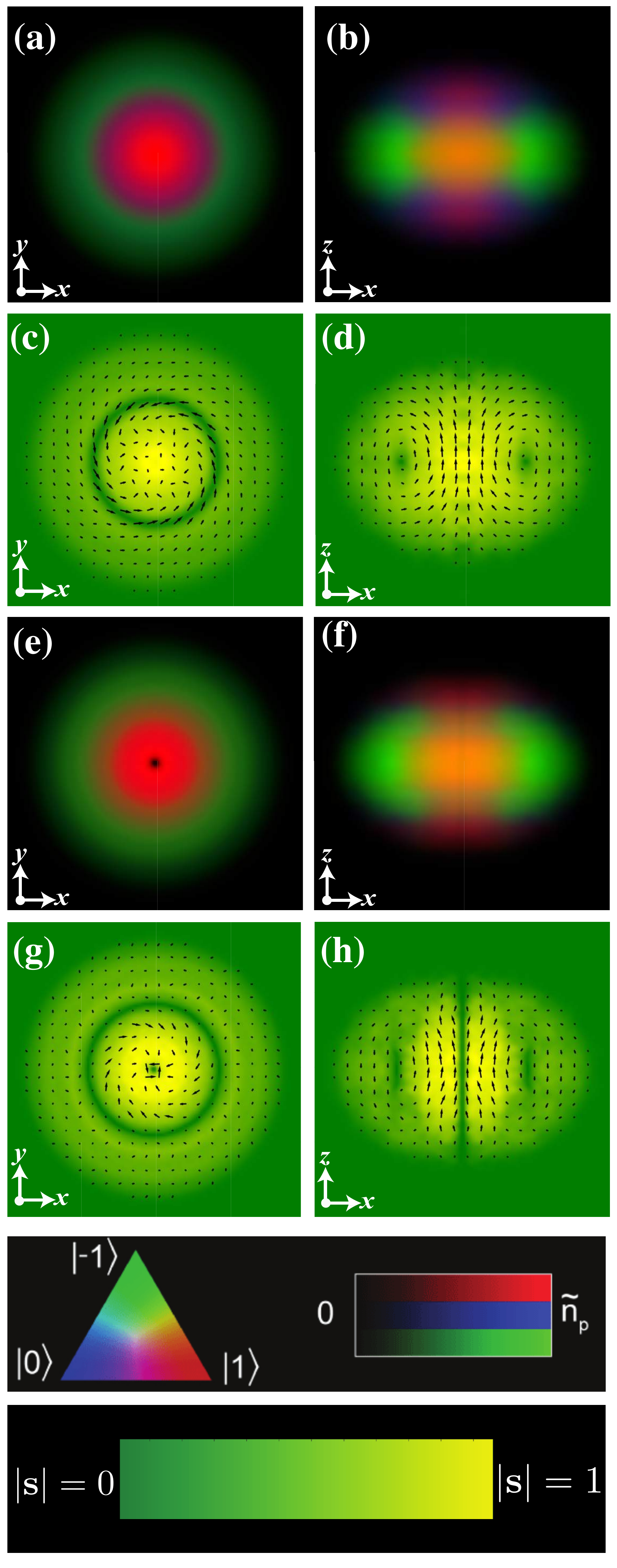}
\caption{\label{fig:5}Decay products of an isolated monopole ($\nemvec=\hat{\ve{r}}$) initially pierced by a straight (a)--(d) singly or (e)--(h) doubly quantized vortex. (a), (b), (e), and (f)~Integrated column densities of the spin components $\ket{\mz}$ in the $z$-quantized basis; the peak column density is $\tilde{n}_{\textrm{p}} = 2.8 \times 10^{11} \textrm{cm}^{-2}$. (c), (d), (g), and (h)~Projections of the spin field $\ve{\spin}(\ve{r})$. The simulations utilize the experimental parameters $g_\mathrm{s}=g_{\mathrm{s}0}$, $\Gamma = \Gamma_0$, and $\omega_z/\omega_r \approx 1.32$. All eight panels correspond to the time $t=340\textrm{ ms}$ and have a field of view of $15.5 \times 15.5\ {\mu}\textrm{m}^2$.}
\end{figure}

Let us take an ideal isolated monopole with $\nemvec=\hat{\ve{r}}$ and pierce it with a singly or doubly quantized straight  $\mathrm{U}(1)$ vortex along the $z$ axis. Such a composite defect corresponds to $\kappa=1$ or $\kappa=2$ in Eq.~\eqref{eq:polar_psi} and induces a nonzero superfluid velocity field $\velocity=-i \hbar \bm{\xi}^\dagger \nabla \bm{\xi}/m$ into the initial state. The resulting states after temporal evolution are depicted in Fig.~\ref{fig:5}. For $\kappa=1$  [Figs.~\ref{fig:5}(a)--\ref{fig:5}(d)], the monopole--vortex composite is found to decay into a coreless spin vortex located along the $z$ axis; the spin texture has an essentially vanishing net magnetization $\ve{M}\approx \ve{0}$ and is reminiscent of the Anderson--Toulouse vortex in superfluid \textsuperscript{3}He-$A$ \cite{And1977.PRL38.508}. Additionally, the coreless spin vortex is encircled by a polar-core vortex ring in the $xy$ plane, similar to the one observed in Ref.~\cite{Pie2009.PRL103.030401} as a decay product of the Dirac monopole configuration. In the case $\kappa=2$  [Figs.~\ref{fig:5}(e)--\ref{fig:5}(h)], the spin texture is similar, except that the vortex along the $z$ axis has a genuinely empty core with vanishing particle density $n=0$. The polar-core vortex ring again appears in the $xy$ plane. 

The nature of the axial vortices in Fig.~\ref{fig:5} can be understood by inspecting the form of the initial spinor $\bm{\xi}_{\textrm{h}}\in G_\mathrm{p}$ [Eq.~\eqref{eq:polar_psi}] and by noticing that the componentwise phase windings remain unchanged during the simulated decay. In the $z$-quantized basis, the monopole accompanied by the singly quantized vortex, $\kappa=1$, corresponds to
\begin{equation}\label{eq:polar_psi_k=1}
\bm{\xi}_{\textrm{h}} =
\frac{1}{\sqrt{2}}\begin{pmatrix}
-\sin\theta\\
\sqrt{2} e^{i \phi} \cos\theta\\
e^{2i \phi} \sin\theta
\end{pmatrix}_\zsub.
\end{equation}
During the decay, the relative populations of the three spinor components change, and the vortex core becomes filled with the windingless  $\mz = 1$ component that corresponds to the local spin magnitude $|\ve{\spin}|=1$. Therefore, we observe an Anderson--Toulouse-type vortex with nonzero local magnetization $n\ve{s}$ at the vortex core [Figs.~\ref{fig:5}(a)--\ref{fig:5}(d)]. For $\kappa=2$, the initial spinor is
\begin{equation}\label{eq:polar_psi_k=2}
\bm{\xi}_{\textrm{h}} =
\frac{1}{\sqrt{2}}
\begin{pmatrix}
-e^{i \phi} \sin\theta\\
\sqrt{2} e^{2i \phi} \cos\theta\\
e^{3i \phi} \sin\theta
\end{pmatrix}_\zsub,
\end{equation}
and thus all three components have nonzero phase windings. Hence, the particle density $n$ should vanish along the vortex core, in agreement with Figs.~\ref{fig:5}(e) and~\ref{fig:5}(f). Finally, for the flowless hedgehog monopole that resulted in the polar-core vortex in Fig.~\ref{fig:4}, the relevant initial spinor is obtained from Eq.~\eqref{eq:polar_psi} by setting $\kappa=0$:
\begin{equation}\label{eq:polar_psi_k=0}
\bm{\xi}_{\textrm{h}} =
-\frac{1}{\sqrt{2}}
\begin{pmatrix}
e^{-i \phi} \sin\theta\\
-\sqrt{2} \cos\theta\\
-e^{i \phi} \sin\theta
\end{pmatrix}_\zsub.
\end{equation}
In this case, the vortex core becomes filled with the windingless  $\mz = 0$ component, resulting in the polar core with $|\ve{\spin}|\approx 0$ and $n\neq 0$. We may similarly explain the appearance of the polar-core vortex in Fig.~\ref{fig:2}. 

\section{\label{sc:discussion}Conclusions}

In summary, we have numerically investigated the evolution of the isolated monopole in a ferromagnetically coupled spin-1 BEC in the absence of any external magnetic fields. Our simulations predict a spontaneous emergence of a polar-core spin vortex in the resulting ferromagnetic order parameter field. We studied both the monopole created according to the previous experiments~\cite{Ray2015.Sci348.544} and an ideal monopole constructed in a spherically symmetric optical potential. Modifying the spin-exchange interaction strength or excluding the  three-body loss does not cause significant qualitative differences in the decay dynamics. Furthermore, we showed that imprinting singly and doubly quantized mass vortices to the initial monopole configuration results in the emergence of quantum vortices of different types. 

Our results provide a fascinating example of dynamical mixing of the polar and ferromagnetic order parameter manifolds, during which the isolated monopole, associated with the nontrivial second homotopy group of the polar phase, becomes transformed into a topological line defect associated with the nontrivial first homotopy group of the ferromagnetic phase. The transition demonstrates that the complex behavior and interconnectedness of the various topological structures supported by the full spin-1 BEC cannot be satisfactorily described by analyzing the system only in terms of the two standard pure manifolds.

\begin{acknowledgments}
This research was supported by the Academy of Finland through its Centres of Excellence Program (Project No.~251748) and Grants No.~135794 and No.~272806. We gratefully acknowledge funding from the Finnish Cultural Foundation (P.K.), the Magnus Ehrnrooth Foundation (K.T. and P.K.), the Emil Aaltonen Foundation (P.K.), and the Technology Industries of Finland Centennial Foundation (P.K.). CSC - IT Center for Science Ltd. (Project No. ay2090) and the Aalto Science-IT project are acknowledged for computational resources. We thank D.~S.~Hall, E.~Ruokokoski, and T.~P.~Simula for insightful discussions.
\end{acknowledgments}
\bibliographystyle{apsrev4-1}
\bibliography{reflist,monopole_evolution_footnotes}

\begin{thebibliography}{59}
\bibitem [{not({\natexlab{a}})}]{note_magnetization_conservation}%
  \BibitemOpen
  \bibinfo {note} {The total magnetization $\ve{M}$ is conserved because the
  $s$-wave scattering only permits $2|\mz=0 \rangle \leftrightharpoons |\mz =
  -1\rangle + |\mz = 1\rangle$ spin-flip processes.}\BibitemShut {Stop}%
\bibitem [{not({\natexlab{b}})}]{note_nematic_order}%
  \BibitemOpen
  \bibinfo {note} {Note, however, that $\bm{\Psi}$ is invariant under the
  transformation $\nemvec \mapsto -\nemvec\ \wedge\ \varphi \mapsto \varphi
  +\pi$ and that both $\nemvec$ and $-\nemvec$ are equivalently valid choices
  for the nematic vector. This is why $G_\mathrm{p}=[S^2\times
  \mathrm{U}(1)]/\mathbb{Z}_2$ and the descriptor of the overall ordering is
  often termed the \emph{nematic axis}.}\BibitemShut {Stop}%
\end{thebibliography}
\end{document}